\documentclass[preprint,12pt]{elsarticle}

\usepackage{amssymb}
\usepackage{doi}
\usepackage{subcaption}
\usepackage{graphicx}
\usepackage{url}
\usepackage{xcolor}
\usepackage{amsmath, amssymb, amsfonts}
\usepackage{multirow}
\usepackage{mathtools}
\usepackage{hyperref}
\usepackage{graphicx}
\usepackage{dsfont}
\usepackage{tikz}
\usepackage{pgfplots}
\usepgfplotslibrary{groupplots}
\newcommand{\onehot}{\operatorname{one\_hot}}
\DeclareMathOperator*{\argmax}{\arg\max}

\journal{Signal Processing}

\begin{document}

\begin{frontmatter}



\title{Clustering of Acoustic Environments with Variational Autoencoders for Hearing Devices} 

\author[label1]{Luan Vin\'icius Fiorio}
\author[label1]{Ivana Nikoloska}
\author[label2]{Wim van Houtum}
\author[label1]{Ronald M. Aarts}

\affiliation[label1]{organization={Eindhoven University of Technology},
            addressline={De Groene Loper, 19},
            city={Eindhoven},
            postcode={5612 AP},
            state={Noord-brabant},
            country={The Netherlands}}
\affiliation[label2]{organization={NXP Semiconductors},
            addressline={High Tech Campus,~60},
            city={Eindhoven},
            postcode={5656 AG},
            state={Noord-Brabant},
            country={Eindhoven}}



\begin{abstract}
Traditional acoustic environment classification relies on: i) classical signal processing algorithms, which are unable to extract meaningful representations of high-dimensional data; or on ii) supervised learning, limited by the availability of labels. Knowing that human-imposed labels do not always reflect the true structure of acoustic scenes, we explore the potential of (unsupervised) clustering of acoustic environments using variational autoencoders (VAEs). We employ a VAE model for categorical latent clustering with a Gumbel-Softmax reparameterization which can operate with a time-context windowing scheme for lower memory requirements, tailored for real-world hearing device scenarios. Additionally, general adaptations on VAE architectures for audio clustering are also proposed. The approaches are validated through the clustering of spoken digits, a simpler task where labels are meaningful, and urban soundscapes, where the recordings present strong overlap in time and frequency. While all variational methods succeeded when clustering spoken digits, only the proposed model achieved effective clustering performance on urban acoustic scenes, given its categorical nature.
\end{abstract}




\end{frontmatter}



\section{Introduction} 

An acoustic environment or soundscape comprise the collection of all acoustic phenomena in a certain space perceived by a listener \cite{iso8253-1_2010}. Noise from acoustic environments are known to not just affect one's well-being \cite{francis2017acoustic}, but also deteriorates one's ability to understand speech (intelligibility), especially in low signal-to-noise ratios with non-stationary signals \cite{wu2024improving}. Hearing devices, such as hearing aids, often take the acoustic environment into account for sound processing, changing its characteristics under different listening conditions \cite{kates1995classification}. Naturally, sound classification algorithms have been developed for recognizing acoustic environments \cite{buchler2005sound}, with most recent approaches based on machine learning \cite{Ting2021,Yellamsetty2021}, using neural networks (NNs).

While most NNs for acoustic environment classification in hearing devices are trained in a supervised manner \cite{Ting2021}, the scarcity of labels has led to the development of semi-supervised learning algorithms \cite{han2016semisupervised}, e.g., pseudo-labeling \cite{Lee2013}, which can generate labels for unknown data based on a few labeled samples \cite{fiorio2023semisupervised}. However, labels are human-imposed categorizations that may not reflect the structure and statistical properties of the data and even be a result of societal bias \cite{shah2025bias}. Moreover, a strong overlap of sound from different sources, in time (simultaneous sound) or in frequency (similar sound), is often present in acoustic environment recordings \cite{barbaro2022linking}. Evidence of such behavior can also be observed by the unoptimal classification accuracy of acoustic scenes obtained through supervised learning \cite{pham2022low}. Although human-designed labels may be of use for case studies or user-focused algorithms \cite{Balling2021Collaboration}, we aim to explore a different context, where we consider labels to be fully missing. This is a step towards a greater reduction of labels in hearing device systems. In that sense, we consider clustering techniques.

Clustering is the process of grouping data into clusters given similarities and characteristics of features \cite{ezugwu2022comprehensive}. Well-established methods like k-means and Gaussian mixture models (GMMs) are still widely applied, but unable to capture complex relations as they are limited to local dependencies in data points, and especially struggle with high-dimensionality data \cite{bishop2006pattern}, which is the case of environmental noise in audio signals. Recently, deep noise suppression embeddings with various (non-variational) autoencoder architectures were analyzed \cite{McLoughlin2025DeepNoiseSuppression}, demonstrating that cluster structures were present in the latent space. The networks considered for the analysis, however, depend on clean speech data as training objective, forming a supervised learning system. Moreover, the study did not have the objective of clustering nor measuring the quality of the obtained clusters by known metrics. An alternative to classical clustering algorithms or supervised approaches lies on deep generative models, which are capable of processing high-dimensionality data and rely on unsupervised learning, being independent of any labels and capable of extracting meaningful information over the data distribution. 

The two most common generative methods modified for (unsupervised) clustering are variational autoencoders (VAEs) \cite{dilokthanakul2017deep} and generative adversarial networks (GANs) \cite{mukherjee2019clusterganlatentspace}. While both perform clustering based on latent variables, VAEs have a more structured latent space \cite{chauhan2018comparative} which directly affects clustering \cite{dilokthanakul2017deep}, and often present a smaller-sized model when compared to GANs. Additionally, GAN models are known for a notorious unstable training \cite{thnahtung2020catastrophic}. Given the hardware-constrained application of hearing devices and the aforementioned characteristics of generative models, we focus on variational autoencoders for the clustering of acoustic environments. VAEs were considered for clustering images and text, usually employing a Gaussian mixture model as their latent space distribution \cite{dilokthanakul2017deep, hershey2016deep, jiang2017variational, ugur2020variational, guo2025gammaclustering}. A generalization was proposed with the use of variational information bottleneck \cite{ugur2020variational}, presenting competitive results. Nevertheless, no generative approach was previously applied specifically to the clustering of audio signals. 

In this work, we demonstrate the potential of variational autoencoders for acoustic environment clustering through a VAE-based clustering model tailored for operation with audio signals\footnote{Parts of the content in this manuscript are present in our previous preprints \cite{fiorio2025unsupervisedvariationalacousticclustering,fiorio2025categoricalunsupervisedvariationalacoustic}. This article is a combination, extension, and thorough revision of those earlier documents, which were not published in a conference nor a journal.}, having the constraints of a hearing device under consideration. Given the discrete nature of clustering, we modify the generative semi-supervised model (M2 model) from \cite{kingma2014semisupervised} for unsupervised clustering through the categorical latent space. For reparameterization and efficient training, the model employs a Gumbel-Softmax \cite{jang2017categorical} function, which also allows for further robustness improvement. Furthermore, in view of the limitations of hearing devices, we devise a sliding window scheme that leverages the Gumbel-Softmax operation for increasing the robustness of the proposed clustering model. Finally, the proposed methods are validated with two tasks: the clustering of spoken digits, representing a scenario where labels are meaningful; and the clustering of urban acoustic environments, where labels are merely an abstraction and the audio content strongly overlaps in time and frequency.

\section{Variational Clustering} 
\label{sec:variational_clustering}
The problem we tackle is the clustering of acoustic environments. More specifically, we consider a dataset $\mathbf{X} = \{\mathbf{x}\}^{N}_{i=1}$ with $N$ independent and identically distributed samples. We aim to cluster $\mathbf{X}$ in $K$ different clusters, using a function $h$ with parameters $\upsilon$, without relying on labels as done in unsupervised learning. As a constraint, the solution $h$ should be sufficiently lightweight to be applied in a hardware-constrained hearing device. Additionally, it should allow for an imperfect availability of environmental recordings, e.g., cases when part of the audio files may be contaminated with foreground-sound speech and cannot be used for soundscape clustering. In the following, we start from the evidence lower bound of variational inference and expand it towards the clustering of audio signals.

\subsection{Continuous variational Inference}
\label{sec:variational_inference}

We consider a model with parameters $\theta$ and a continuous latent variable $\mathbf{z}$ representing the underlying data distribution. The model's ability to represent data can be measured by its likelihood $p_\theta(\mathbf{x}|\mathbf{z})$, since it tells us how probable it is to obtain $\mathbf{x}$ given parameters $\theta$. If $\mathbf{z}$ was known, we could find the model by directly maximizing $p_\theta(\mathbf{x}|\mathbf{z})$ with respect to $\theta$. As $\mathbf{z}$ is an unobserved (latent) variable, we instead aim to maximize the marginal likelihood, $p_{\theta}(\mathbf{x})$. However, $p_{\theta}(\mathbf{x})$ is intractable for the majority of real-world problems given its intractable posterior $p_\theta(\mathbf{z}|\mathbf{x})$ -- which relies on knowledge over $\mathbf{z}$. Therefore, we introduce a \textit{variational} distribution with parameters $\phi$ to approximate the posterior $q_\phi(\mathbf{z}|\mathbf{x}) \approx p_\theta(\mathbf{z}|\mathbf{x})$. With $q_\phi(\mathbf{z}|\mathbf{x})$, we can find the \textit{variational lower bound} $\mathcal{L}$, also called the evidence lower bound (ELBO) \cite{kingma2014autoencoding}:
\begin{multline}
\label{eq:variationallowerbound}
    \log p_{\theta}(\mathbf{x}) \geq \mathbb{E}_{q_\phi(\mathbf{z}|\mathbf{x})} \left[ \log p_{\theta}(\mathbf{x}|\mathbf{z}) + \log \frac{p_{\theta}(\mathbf{z})}{q_\phi(\mathbf{z}|\mathbf{x})} \right] \\
    = \mathbb{E}_{q_\phi(\mathbf{z}|\mathbf{x})} \left[ \log p_{\theta}(\mathbf{x}|\mathbf{z}) \right] - D_{KL} ( q_\phi(\mathbf{z}|\mathbf{x}) \, || \, p_{\theta}(\mathbf{z}))
    = \mathcal{L}(\theta, \phi).
\end{multline}

In \eqref{eq:variationallowerbound}, $\mathbb{E}_{q_\phi(\mathbf{z}|\mathbf{x})} \left[ \log p_{\theta}(\mathbf{x}|\mathbf{z}) \right]$ is the expectation over the distribution $q_\phi(\mathbf{z}|\mathbf{x})$ of $\log p_{\theta}(\mathbf{x}|\mathbf{z})$, and represents the reconstruction error of $\mathbf{x}$ from $\mathbf{z}$, i.e., how well the latent variables $\mathbf{z}$ explain the data $\mathbf{x}$. The $D_{KL}$ term stands for the Kullback-Leibler (KL) divergence, which quantifies the difference between the prior $p_{\theta}(\mathbf{z})$ and the variational distribution $q_\phi(\mathbf{z}|\mathbf{x})$. This can be seen as a regularization to make the variational distribution closer to the desired/known prior distribution of the latent variables. The objective of variational inference is to maximize the ELBO in \eqref{eq:variationallowerbound} by optimizing $\theta$ and $\phi$. This corresponds to learning a good approximation $q_\phi(\mathbf{z}|\mathbf{x})$ to the true posterior $p_\theta(\mathbf{z}|\mathbf{x})$, while also optimizing the generative model $p_\theta(\mathbf{z}|\mathbf{x})$.

\begin{figure}[!t]
\centering
    \begin{subfigure}{.99\textwidth}
        \centering
        \includegraphics[width=0.5\textwidth]{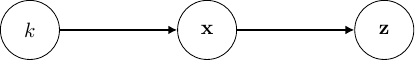}  
        \caption{Inference model}
        \label{fig:inferencemodel}
        \vspace{5mm}
    \end{subfigure}
    \begin{subfigure}{.99\textwidth}
        \centering
        \includegraphics[width=0.5\textwidth]{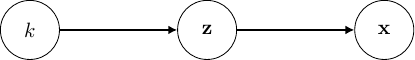}
        \caption{Generative model}
        \label{fig:generativemodel}
    \end{subfigure}
\caption{Inference/generative model for clustering with a continuous latent variable $\mathbf{z}$.}
\label{fig:models}
\vspace{-5mm}
\end{figure}

\subsection{Extension to clustering}
\label{ssec:extension_clustering}

Given the complexity of the problem, we consider the VAE \cite{kingma2014autoencoding} framework because of its capacity to learn the underlying distribution of data through a structured latent \cite{dilokthanakul2017deep}, which can be leveraged for clustering \cite{chauhan2018comparative} by grouping data based on its probabilistic behavior. The VAE is composed by a neural network encoder, which takes the data $\mathbf{x}$ to a latent representation $\mathbf{z}$, and a NN decoder that generates data based on the latent variables. 

To allow clustering behavior, we choose a multivariate Gaussian mixture model (GMM) prior $p_\varphi(\mathbf{z})$, replacing $p_\theta(\mathbf{z})$ in \eqref{eq:variationallowerbound}, given by
\begin{equation}
\label{eq:gmmprior}
    p_\varphi(\mathbf{z}) = \sum_{k} \pi_k \ \mathcal{N}(\mathbf{z}; \boldsymbol{\mu}_k, \boldsymbol{\Sigma}_k), 
\end{equation}
where $k$ is one component in the mixture, i.e., a cluster. The multivariate GMM defined in \eqref{eq:gmmprior} is based on multivariate Gaussian $\mathcal{N}$ with means $\boldsymbol{\mu}_k$ and variances $\boldsymbol{\Sigma}_k$, whose dimension is $d_{\boldsymbol{\mu}_k} = d_{\boldsymbol{\Sigma}_k} = d_\mathbf{z}$, linearly combined by a weighting vector $\pi_k$. The GMM parameters are condensed in a variable $\varphi \coloneqq \{\pi_k, \boldsymbol{\mu}_k, \boldsymbol{\Sigma}_k\}_{k=1}^{K}$, for $K$ components. 

As shown in Fig.~\ref{fig:models}, for inference, the data are generated from a component $k$ of an unknown GMM $q_\phi(\mathbf{x}|k)$. In addition, the latent variables have a GMM prior $p_\varphi(\mathbf{z})$. All parameters of the model ($\theta$, $\phi$, and $\varphi$) are optimized simultaneously to maximize the ELBO, modified from \eqref{eq:variationallowerbound} to meet the GMM prior as
\begin{equation}
\label{eq:variationallowerboundclustering}
    \mathcal{L}(\theta, \phi, \varphi) =
    \mathbb{E}_{q_\phi(\mathbf{z}|\mathbf{x})} \left[ \mathbf{a}_{\mathbf{x}} \log p_{\theta}(\mathbf{x}|\mathbf{z}) \right] - D_{KL} ( q_\phi(\mathbf{z}|\mathbf{x}) \, || \, p_\varphi(\mathbf{z})).
\end{equation}
Notice that, we explicitly include the term $\mathbf{a}_{\mathbf{x}}$, which is sound activity mask for $\mathbf{x}$ over time. Specifically for audio processing, ${a}_{\mathbf{x},t}=1$ when time step $t$ contains sound activity, and ${a}_{\mathbf{x},t}=0$ for a silent time step. This is necessary to avoid the clustering of silence/zero padding patterns. The ELBO in \eqref{eq:variationallowerboundclustering}, except from the extension of the sound activity mask, is equivalent to the variational information bottleneck
with Gaussian mixture model (VIB-GMM) proposed by \cite{ugur2020variational}. 

Importantly, when using backpropagation, the cost function based on (8) can be written as
\begin{equation}
    \min_{\theta, \phi, \varphi} \left( -\frac{1}{N} \sum_{i=1}^N \mathcal{L}^{(i)}(\theta, \phi, \varphi) \right),
\end{equation}
where $\mathcal{L}^{(i)}$ denotes the ELBO for the $i$-th data sample, and the negative sign reflects that we minimize the negative ELBO during training.

Nevertheless, in practice, we cannot directly sample $\mathbf{z} \sim p_\phi(\mathbf{z})$ to compute the reconstruction error in (8), since sampling is a non-differentiable operation. To enable gradient-based optimization, we apply the reparameterization trick \cite{kingma2014autoencoding} combined with Monte Carlo sampling \cite{ugur2020variational},
\begin{equation}
\label{eq:carlo}
    \mathbb{E}_{q_\phi(\mathbf{z}|\mathbf{x})} [\log p_\theta(\mathbf{x}|\mathbf{z})] \approx \frac{1}{M} \sum_{m=1}^M \log p_\theta(\mathbf{x}|\mathbf{z}^{(m)}),
\end{equation}
where each sample $\mathbf{z}^{(m)}$ is obtained via
\begin{equation}
    \mathbf{z}^{(m)} = \boldsymbol{\mu}_\phi + \boldsymbol{\Sigma}_\phi^{1/2} \cdot \boldsymbol{\epsilon}^{(m)}, \quad \boldsymbol{\epsilon}^{(m)} \sim \mathcal{N}(0, I).
\end{equation}

The KL divergence term in (8), which measures the divergence between a single-component multivariate Gaussian and a Gaussian mixture model (GMM) with $C$ components, does not have a closed-form solution. However, it can be approximated under the assumption that both covariance matrices are diagonal,
\begin{equation}
    \Sigma_\phi = \text{diag}(\{\Sigma_{\phi,j}\}_{j=1}^{d_z}), \quad \Sigma_k = \text{diag}(\{\Sigma_{k,j}\}_{j=1}^{d_z}),
\end{equation}
resulting in
\begin{multline}
\label{eq:xunxo_kl}
    D_{KL} ( q_\phi(\mathbf{z}|\mathbf{x}) \, || \, p_\varphi(\mathbf{z})) \approx \\
    -\log \sum_{k} \pi_k \exp \Biggl( -\frac{1}{2} \sum_{j} \Biggl[ \frac{(\mu_{\phi,j} - \mu_{k,j})^2}{\Sigma_{k,j}} 
    + \log \left( \frac{\Sigma_{k,j}}{\Sigma_{\phi,j}} \right) 
    - 1 + \frac{\Sigma_{\phi,j}}{\Sigma_{k,j}} \Biggr] \Biggr).
\end{multline}
Moreover, the cluster assignment $k$ is determined by the component with the highest posterior probability:
\begin{equation}
\label{eq:pc_gmm}
    q_\phi(k|\mathbf{x}) \approx p_\phi(k|\mathbf{z}) = \frac{\pi_k \  \mathcal{N}(\mathbf{z}; \boldsymbol{\mu}_k, \boldsymbol{\Sigma}_k)}{\sum_{k'} \pi_{k'} \ \mathcal{N}(\mathbf{z}; \boldsymbol{\mu}_{k'}, \boldsymbol{\Sigma}_{k'})},
\end{equation}
\begin{equation}
\label{eq:k_argmax}
    k = \argmax_k \, p_\phi(k|\mathbf{z}).
\end{equation}

For hearing devices, the data $\mathbf{X}$ is composed of audio recordings, which strongly overlap in time and frequency. A continuous latent variable might not be sufficient for enforcing separate and compact clusters. Categorical latent variables, on another hand, enforce the posterior for each data point to be concentrated on one of $K$ bins, which is, naturally, close to the task of clustering. Next, we extend variational clustering to include a categorical latent variable.

\section{Continuous-Categorical Variational Clustering}

\label{sec:CC_variational_inference}
\begin{figure}[!t]
\centering
    \begin{subfigure}{.99\textwidth}
        \centering
        \includegraphics[width=0.5\textwidth]{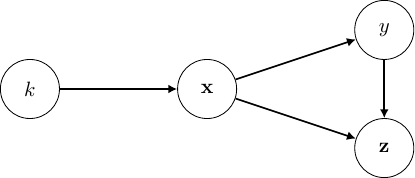}  
        \caption{Inference model}
        \label{fig:inferencemodel_m2}
        \vspace{5mm}
    \end{subfigure}
    \begin{subfigure}{.99\textwidth}
        \centering
        \includegraphics[width=0.5\textwidth]{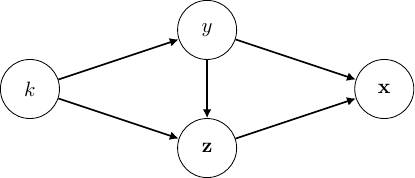}
        \caption{Generative model}
        \label{fig:generativemodel_m2}
    \end{subfigure}
\caption{Inference/generative model for clustering with a categorical ($y$) and a continuous ($\mathbf{z}$) latent variable.}
\label{fig:models_m2}
\vspace{-5mm}
\end{figure}

The inclusion of a categorical latent variable in the inference and generative models, as presented in Fig.~\ref{fig:models_m2}, is similar to the M2 architecture\footnote{Notice that Rui Shu hinted, in his blog, that the M2 model could be a natural choice for clustering \cite{shu2016gmvae}.} \cite{kingma2014semisupervised} proposed for semi-supervised learning. Here, data $\mathbf{x}$ is assumed to come from an unknown discrete cluster $k$. In the continuous-categorical case, each observation $\mathbf{x}$ is considered to be generated by a class $k$ of a categorical latent variable $y$ and a continuous latent variable $\mathbf{z}$. The joint distribution is given by
\begin{equation}
    p_\theta(\mathbf{x}, y, \mathbf{z}) = p_\theta(\mathbf{x}|y,\mathbf{z}) p_\theta(y) p_\theta(\mathbf{z}).
\label{eq:generative_model}
\end{equation}

We are interested in the true posterior $p_\theta(y, \mathbf{z} | \mathbf{x})$, as it tells us how likely each latent configuration is given an observed data point. However, given the intractability of $p_\theta(y, \mathbf{z} | \mathbf{x})$, we define a \emph{variational} distribution with parameters $\upsilon$ and $\phi$,
\begin{equation}
    q_{\upsilon,\phi}(y, \mathbf{z} | \mathbf{x}) = q_\phi(\mathbf{z}|\mathbf{x}, y) q_\upsilon(y|\mathbf{x}),
\label{eq:inference_model}
\end{equation}
such that $ q_{\upsilon,\phi}(y, \mathbf{z} | \mathbf{x}) \approx p_\theta(y, \mathbf{z} | \mathbf{x})$. Therefore, we can find the ELBO for the continuous-categorical case as
\begin{multline}
    \mathcal{L}(\theta, \upsilon, \phi) = \mathbb{E}_{q_{\upsilon,\phi}(y, \mathbf{z} | \mathbf{x})}[ \mathbf{a}_{\mathbf{x}} \log  p_\theta(\mathbf{x}|y,\mathbf{z})] \\
    - \beta(D_{KL}(q_\phi(\mathbf{z}|y, \mathbf{x}) || p_\theta(\mathbf{z})) + D_{KL}(q_\upsilon(y|\mathbf{x})||p_\theta(y))).
\label{eq:elbo}    
\end{multline}

In \eqref{eq:elbo}, $\mathbb{E}_{q_{\upsilon,\phi}(y, \mathbf{z} | \mathbf{x})}[ \mathbf{a}_{\mathbf{x}} \log  p_\theta(\mathbf{x}|y,\mathbf{z})]$ is the expectation of $\mathbf{a}_{\mathbf{x}} \log  p_\theta(\mathbf{x}|y,\mathbf{z})$ over $q_{\upsilon,\phi}(y, \mathbf{z} | \mathbf{x})$, representing the reconstruction error. The KL divergence terms $D_{KL}(q_\phi(\mathbf{z}|y, \mathbf{x}) || p_\theta(\mathbf{z}))$ and $D_{KL}(q_\upsilon(y|\mathbf{x})||p_\theta(y))$ respectively, for the continuous and the categorical encoders, serve as regularization by enforcing their output distribution to remain close to the chosen priors. As previously done for \eqref{eq:variationallowerboundclustering}, a sound activity mask of $\mathbf{x}$ over time, $\mathbf{a}_{\mathbf{x}}$, is added to the reconstruction error, such that the clustering is focused on actual sound parts, not silence or zero padding. Additionally, a non-trainable hyperparameter $\beta$ is included for smoother training \cite{higgins2017betavae} and should be tuned based on data.

For the continuous-categorical case, the clusters can be chosen based on the maximum probability of data point $\mathbf{x}$ belonging to the $k$th cluster $q_\upsilon(k|\mathbf{x}) = p_\theta(y|k)$, obtained as
\begin{equation}
    k = \argmax_{n} \ \{y_n\}_{n=1}^{d_y},
\end{equation}
where $d_y = K$ is the dimension of $y$.

While the continuous latent variable $\mathbf{z}$ can be sampled by the reparameterization trick, as mentioned in Section~\ref{sec:variational_clustering}, the same trick cannot be applied to a categorical distribution. We, thus, consider a Gumbel-Softmax function \cite{jang2017categorical} to sample from a categorical distribution and have a smooth training, also allowing for the application of the time-context windowing described in Section~\ref{sec:windowing}. We describe the Gumbel-Softmax distribution in the following.

\subsection{Sampling from a categorical distribution}

The foundation of this approach is the Gumbel-Max trick \cite{maddison2015asampling}, which allows sampling from a categorical distribution using a deterministic transformation combined with noise. Consider a categorical distribution over $K$ classes with probabilities $[\pi_1, \ldots, \pi_K]$. The Gumbel-Max trick samples from this distribution by computing
\begin{equation}
    {y}_{\mathds{1}} = \onehot \left( \argmax_k \ [\log \pi_k + g_k] \right) ,
\label{eq:gumbel_max}        
\end{equation}
where each $g_k \sim \text{Gumbel}(0, 1)$ is generated via
\begin{equation}
    g_k = -\log(-\log(u_k)), \quad u_k \sim \text{Uniform}(0, 1),
\end{equation}
transforming the sampling process into a maximization over noisy logits. Nevertheless, the $\argmax$ operation is not differentiable, posing a problem for gradient-based optimization.

To address it, we replace the argmax with a differentiable approximation: the softmax function, which leads to the Gumbel-Softmax (GS) distribution \cite{jang2017categorical}:
\begin{equation}
    y = \frac{\exp \left( (\log \pi_k + g_k)/\tau \right)}{\sum_{j=1}^{K} \exp \left( (\log \pi_j + g_j)/\tau \right)}, \quad \text{for } k = 1,\text{...}, K.
\label{eq:gumbel_softmax}    
\end{equation}
Here, $\tau$ is a temperature parameter that controls the smoothness of the distribution. As $\tau \to 0$, the GS approaches a categorical distribution (i.e., becomes more discrete). For higher $\tau$, the distribution becomes smoother and more continuous.

\begin{figure}[!t]
    \centering
    \begin{tikzpicture}
    \hspace{-3mm} 
      \begin{groupplot}[
          group style={
            group size=2 by 2,
            horizontal sep=1.75cm,
            vertical sep=1.4cm,
          },
          scale=0.7,
          width=4.75cm,
          height=4cm,
          xlabel={$k$},
          xtick={1,10},
          ylabel={$y$},
          xlabel style={yshift=1.5mm},
          title style={yshift=-1.5mm},
          ybar,
          grid=both,
          grid style={line width=.1pt, draw=gray!20},
          major grid style={line width=.2pt,draw=gray!50},
      ]
      
      \nextgroupplot[
          title={\textcolor{black}{$\tau = 0.01$}},
          ymin=0, ymax=1.1,
      ]
      \addplot[fill=black!50!white, bar width=6pt] coordinates {
          (1,0.0000) (2,0.0000) (3,0.0000) (4,0.0000) (5,3.1838e-32)
          (6,1.0000) (7,1.1245e-35) (8,0.0000) (9,0.0000) (10,1.2752e-43)
      };
    
      \nextgroupplot[
          title={\textcolor{black}{$\tau = 0.5$}},
          ymin=0, ymax=0.55,
          ]
      \addplot[fill=black!50!white, bar width=6pt] coordinates {
          (1,0.0349) (2,0.0203) (3,0.0119) (4,0.0604) (5,0.1218)
          (6,0.5193) (7,0.1039) (8,0.0532) (9,0.0023) (10,0.0720)
      };
    
      \nextgroupplot[
          title={\textcolor{black}{$\tau = 1.0$}},
          ymin=0, ymax=0.30,
      ]
      \addplot[fill=black!50!white, bar width=6pt] coordinates {
          (1,0.0712) (2,0.0543) (3,0.0415) (4,0.0937) (5,0.1330)
          (6,0.2747) (7,0.1228) (8,0.0880) (9,0.0185) (10,0.1023)
      };
    
      \nextgroupplot[
          title={\textcolor{black}{$\tau = 100.0$}},
          ymin=0, ymax=0.11,
          ytick={0, 0.05, 0.1}, 
          yticklabels={0, 0.05, 0.1},
      ]
      \addplot[fill=black!50!white, bar width=6pt] coordinates {
          (1,0.0999) (2,0.0996) (3,0.0993) (4,0.1002) (5,0.1005)
          (6,0.1012) (7,0.1004) (8,0.1001) (9,0.0985) (10,0.1002)
      };
    
      \end{groupplot}
    \end{tikzpicture}
    \vspace{-2mm}
    \caption{Gumbel-Softmax distribution example plot for 10 classes with different values of $\tau$.}
    \label{fig:gumbel_plot}
    \vspace{-5mm}
\end{figure}
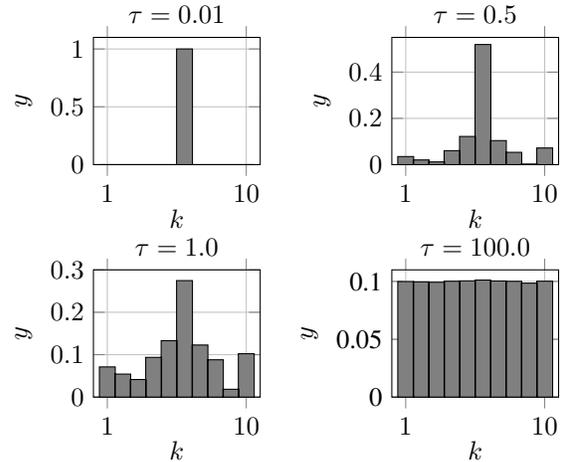

To illustrate, Fig.~\ref{fig:gumbel_plot} shows the GS distribution plot for 10 classes, with different values of $\tau$. The softmax temperature $\tau$ is specially interesting when the GS distribution is used for clustering, as a smaller $\tau$ results in more distinct and dense clusters. Therefore, as a natural choice, we choose to cluster over $y$ with a monotonic reduction of $\tau$ over training. Next, we propose a time-context windowing scheme for the continuous-categorical model, such that its operation in inference mode can be applied for real-world hearing devices.

\section{Time-Context Windowing}
\label{sec:windowing}

In practical hearing devices, a voice activity detection mask \cite{sehgal2018vad} is often considered for detecting traces of speech in the captured audio. To allow for a clear clustering of acoustic environment, we assume a device where a voice-activity mask is used for separating parts of the signal where only background sound is present, as done in \cite{lee2020dynamicnoiseembeddingnoise}. Therefore, only a subset of a complete audio recording can be used for clustering the soundscape. 

\begin{figure}[!t]
    \centering
    \includegraphics[width=0.6\linewidth]{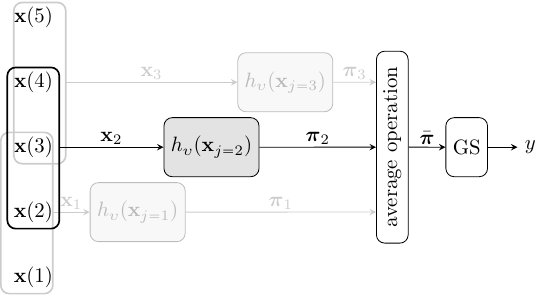}
    \caption{Example of sliding window clustering inference for $w=3$, hop of 1 sample, with $N=5$ samples. $\mathbf{x}(n)$ is the $n$-th time bin of an input sample $\mathbf{x}^{(i)}$, $\mathbf{x}_{j}$ is the input window $j$ with estimated logits $\boldsymbol{\pi}_{j}$, $\bar{\boldsymbol{\pi}}$ are the average logits, $\mathrm{GS}$ is the Gumbel-Softmax function, and $y$ are the cluster probabilities.}
    \label{fig:uvacwindow}
    \vspace{-5mm}
\end{figure}

We propose a windowing scheme, described in the following, that allows a categorical-based clustering model to cluster the environment without access to a complete background-only audio trace. As shown in Figure~\ref{fig:uvacwindow}, the sliding window operates in inference mode and consists of sliding through the time dimension of the input, combining the logits $\boldsymbol{\pi}$ over the whole sequence for a more reliable estimation of cluster probabilities through the GS function. This extension is based on a sliding window approach previously applied in speech enhancement \cite{fiorio2024spectral} and communications \cite{Karanov2019EndToEnd}.


Given the spectrogram dataset $\mathbf{X}$, each sample $\mathbf{x}$ consists of a sequence of length $T$ (representing the time dimension), with $F$ frequency bins per time step. Our method applies a sliding window of length $w$ across the $i$-th sample vector $\mathbf{x}^{(i)} = [x(1), x(2), \dots, x(T)]$. For each window position $j$, we extracts a segment
\begin{equation}
    \mathbf{x}_j = [\mathbf{x}(j), \mathbf{x}(j+1), \dots, \mathbf{x}(j + w - 1)].
\end{equation}
This windowed segment is fed into a neural network model $h_\upsilon$, which outputs the corresponding logits
$h_\upsilon(\mathbf{x}_j) = \boldsymbol{\pi}_j$. All logits from the $N_w$ windows are then averaged as
\begin{equation}
    \bar{\boldsymbol{\pi}} = \frac{1}{N_w} \sum_{j=1}^{N_w} \boldsymbol{\pi}_j.
\end{equation}
Finally, the averaged logits $\bar{\boldsymbol{\pi}}$ are passed through the GS function to produce the output probabilities $y$. 

It is worth mentioning that the inference overlap between windows in inference can be chosen based on a trade-off of computational complexity and performance. A smaller hop results in better performance with the cost of elevated computational complexity. For this work, we utilize a hop of 50\% the size of the window.

\section{Model architectures}
\label{sec:architecture}

We consider two different variational autoencoder architectures for the application of unsupervised audio clustering. Each architecture is aimed at the optimization of the ELBOs derived in Sections~\ref{sec:variational_clustering} and \ref{sec:CC_variational_inference}, and is described as follows.

\subsection{VAE with continuous prior}

\begin{figure}[!t]
    \centering
    \includegraphics[width=0.5\linewidth]{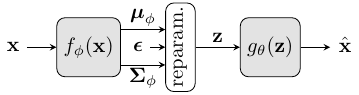}
    \caption{Diagram of the neural network-based VAE clustering model with a (continuous) Gaussian mixture model prior. $q_\phi(\mathbf{z}|\mathbf{x})$ is represented by a NN  $f_\phi(\mathbf{x}) = [\boldsymbol{\mu}_\phi, \boldsymbol{\Sigma}_\phi]$. The decoder $p_\theta(\mathbf{x}|\mathbf{z})$ is a NN $g_\theta(\mathbf{z}) = [\hat{\mathbf{x}}]$.}
    \label{fig:vae-nn-model}
    \vspace{-4mm}
\end{figure}

Fig.~\ref{fig:vae-nn-model} illustrates the neural network-based variational autoencoder model used for clustering, which assumes a continuous prior over the latent space -- in this case, a GMM. The \textit{inference model}, shown in Fig.~\ref{fig:inferencemodel}, is represented by the encoder $q_\phi(\mathbf{z}|\mathbf{x}) = \mathcal{N}(\mathbf{z}; \boldsymbol{\mu}_\phi, \boldsymbol{\Sigma}_\phi)$. We implement the encoder as a neural network $f$ with parameters $\phi$ and input $\mathbf{x}$, producing the multivariate output $f_\phi(\mathbf{x}) = [\boldsymbol{\mu}_\phi, \boldsymbol{\Sigma}_\phi]$. Similarly, the \textit{generative model}, shown in Fig.~\ref{fig:generativemodel}, is represented by the decoder $p_\theta(\mathbf{x}|\mathbf{z}) = \hat{\mathbf{x}} = g_\theta(\mathbf{z})$. Here, $g$ is a neural network with parameters $\theta$ that takes the latent variable $\mathbf{z}$ as input and generates the reconstructed data $\hat{\mathbf{x}}$. For the inference mode, however, the generative part of the model is not needed, as we are only interested in clustering in the latent space. Therefore, only the encoder $f_\phi(\mathbf{x})$ is necessary in inference time, with clustering decision taken with \eqref{eq:pc_gmm} and \eqref{eq:k_argmax}. 

Moreover, generative models tend to require a massive number of parameters to achieve desirable performance \cite{kaplan2020scalinglawsneurallanguage}. Thus, we define a reduced-size convolutional-recurrent variational autoencoder composed of a convolutional-recurrent encoder $q_\phi (\mathbf{z}|\mathbf{x})$, a latent space with prior $p_\varphi(\mathbf{z})$, and a (mirrored) convolutional decoder $p_\theta(\mathbf{x}|\mathbf{z})$. We define the VAE with continuous prior as the VIB-GMM model, the same used in \cite{jiang2017variational,ugur2020variational}, and its audio clustering (AC) version with a convolutional-recurrent structure, the VIB-GMM-AC, which detailed schematic can be seen in Fig.~\ref{fig:VAE}.

During training considering the time-context windowing scheme, the recurrent layer's final hidden state at window $j-1$ is carried as initial state at window $j$, but the gradients of the carried state are detached for training efficiency. For the first window of each data sample, the initial state is initialized as zero. The training losses, calculated with \eqref{eq:elbo}, are obtained per-window and combined at the end of each batch of samples, making sure that the contextual relation between windows in a data sample is taken into account. Importantly, no overlap between windows is considered in training to avoid overfitting. Furthermore, training is executed with full knowledge of the acoustic environment recording, taking entire files composed of background noise only. We also consider pure-noise files for evaluation. Notice that, in practice, voice activity detectors are used to filter-out audio traces where speech dominates the signal, and can be estimated in different ways \cite{lee2020dynamicnoiseembeddingnoise}. Its use is out of scope for this work and we cluster acoustic environment audio directly. Next, we propose variational autoencoder architectures for the context-aware clustering of audio.


\subsection{VAE with continuous-categorical prior}

\begin{figure}[!t]
    \centering
    \includegraphics[width=0.7\linewidth]{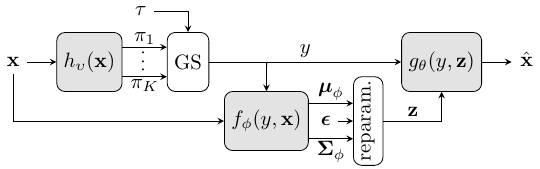}
    \caption{Diagram of the continuous-categorical neural network-based VAE model for clustering considering both continuous ($\mathbf{z}$) and a categorical ($y$) latent variable. $q_\upsilon(y|\mathbf{x})$ is represented by a NN $h_\upsilon(\mathbf{x})=[\boldsymbol{\pi}_k]$ and $\tau$ is the Gumbel-Softmax temperature. $f_\phi(y,\mathbf{x})=[\boldsymbol{\mu}_\phi, \boldsymbol{\Sigma}_\phi]$ is a NN equivalent to $q_\phi(\mathbf{z}|y,\mathbf{x})$. The decoder $p_\theta(\mathbf{x}|y,\mathbf{z})$ is a NN $g_\theta(y,\mathbf{z}) = [\hat{\mathbf{x}}]$.}
    \label{fig:M2_model}
    \vspace{-2mm}
\end{figure}

The neural network-based application of the continuous-categorical clustering model is shown in Fig.~\ref{fig:M2_model} and follows the inference and generative models from Fig.~\ref{fig:models_m2}. For the discriminative inference encoder $q_\upsilon(y|\mathbf{x})$, we use a neural network denoted by $h$ with parameters $\upsilon$, which outputs $K$ (soft) probabilities $h_\upsilon(\mathbf{x}) = [\pi_1, \ldots, \pi_K]$. The inference encoder $q_\phi(\mathbf{z}|y, \mathbf{x})$ is represented by another NN, denoted by $f$ with parameters $\phi$, such that $f_\phi(y, \mathbf{x}) = [\boldsymbol{\mu}_\phi, \boldsymbol{\Sigma}_\phi]$, where $\boldsymbol{\mu}_\phi$ and $\boldsymbol{\Sigma}_\phi$ are used to sample $\mathbf{z}$ via the re-parameterization trick \cite{kingma2014autoencoding} $\mathbf{z} = \boldsymbol{\mu}_\phi + \boldsymbol{\Sigma}_\phi^{1/2} \cdot \boldsymbol{\epsilon}$, with $\boldsymbol{\epsilon} \sim \mathcal{N}(0, I)$. Furthermore, the generative decoder is modeled by a NN $g$ with parameters $\theta$, which reconstructs the data $\mathbf{x}$ as $g_\theta(y, \mathbf{z}) = [\hat{\mathbf{x}}]$. Importantly, during inference time for clustering, only $h_\upsilon(\mathbf{x})$ and the GS distribution \eqref{eq:gumbel_softmax} with hard outputs are required.

\begin{figure}[!t]
\centering
    \begin{subfigure}{.99\textwidth}
        \centering
        \includegraphics[width=0.815\textwidth]{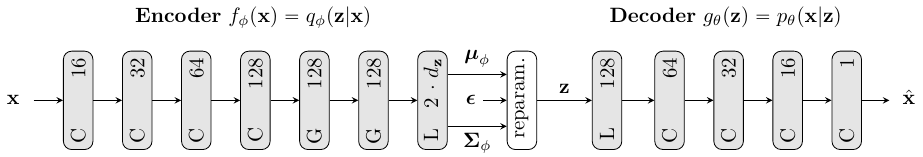}
        \caption{VIB-GMM-AC architecture (similarly for VaDE-AC)}
        \label{fig:VAE}
    \end{subfigure} \\ \vspace{3mm}
    \begin{subfigure}{.99\textwidth}
        \centering
        \includegraphics[width=0.9\textwidth]{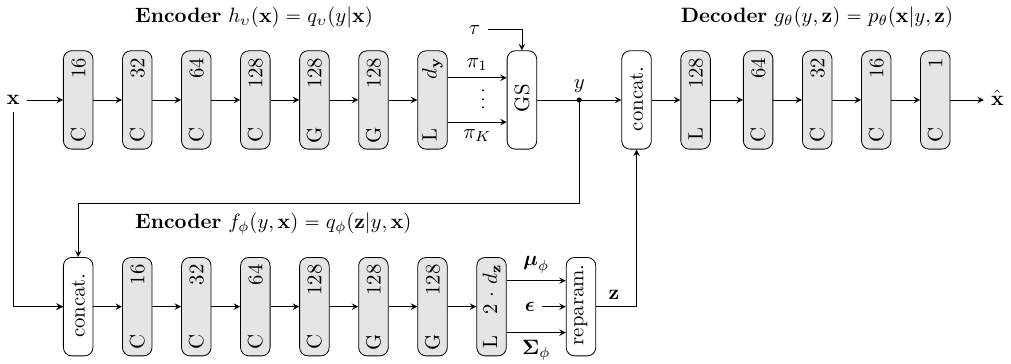}
        \caption{M2-AC architecture}
        \label{fig:VAE-M2}
    \end{subfigure}
\caption{Proposed VAE architectures for continuous prior (a) and continuous-categorical latent spaces (b). The number in each layer indicates output channels. The $\mathrm{C}$ encoder layers consist of Conv2D, while $\mathrm{G}$ layers are GRUs. The decoder layers are tranposed convolutions Conv2D.T. All $\mathrm{C}$ kernels are (8,8) with stride (2,2) and padding (3,3). Every $\mathrm{C}$ layer is followed by 2D batch normalization and a ReLU activation, except the last convolution before a GRU layer, which employs a tanh, and the last convolution of the decoder, followed by a sigmoid function. Linear layers ($\mathrm{L}$) have no activation. ``concat'' is a concatenation in the channel (latent) dimension.}
\label{fig:architectures}
\vspace{-4mm}
\end{figure}

The architecture considered for the audio clustering M2 model follows similar characteristics to the VIB-GMM-AC, nonetheless, including one additional encoder and minor changes, which can be seen in Fig.~\ref{fig:VAE-M2}. Both inference networks $h_\upsilon = q_\upsilon(y|\mathbf{x})$ and $f_\phi(y,\mathbf{x}) = q_\phi(\mathbf{z}|y,\mathbf{x})$ are convolutional-recurrent layers. A linear layer is used in model $h$ to output $K$ values $\pi_k$, dimension $d_y$, which are used as input to the GS distribution. Model $f$ also uses a linear layer for latent dimension conversion, as explained for the VIB-GMM-AC. The outputs of both GS and continuous reparameterization are concatenated and input to the decoder, $g_\theta(y,\mathbf{z}) = p_\theta(\mathbf{x}|y,\mathbf{z})$, which is a mirrored version of the encoder. Hereafter, we refer to this model as M2-AC: M2 model extended for audio clustering. The detailed schematics of the M2-AC model is shown in Fig.~\ref{fig:VAE-M2}. Similarly as before, all layers are initialized using Kaiming uniform initialization for the weights, and all biases are set to zero. 

\subsection{Practical considerations}
\label{ssec:practical}

We identified several key implementation aspects in the continuous-categorical audio clustering settings that significantly impact performance, supported by ablation studies presented in \ref{appendix:ablation_studies}. The main conclusions are described in the following.

\begin{enumerate}
    \item Audio signals exhibit strong time-context dependencies, especially in longer sequences. Incorporating recurrent layers in the encoder architecture substantially improves performance. In contrast, omitting them can lead to unstable training behavior, including loss collapse;
    \item Both the encoder and decoder need two-dimensional convolutions. We found that only applying convolutions along the frequency dimension -- combined with global time pooling -- resulted in poor clustering performance. This highlights the importance of preserving both time and frequency structure;
    \item When input signals are zero-padded to meet a fixed duration or contain irrelevant silence, a time mask should be applied during reconstruction error computation. This ensures that non-informative regions do not negatively affect the model’s learning. Refer to the time mask $\mathbf{a}_x$ in the ELBO formulations \eqref{eq:variationallowerboundclustering} and \eqref{eq:elbo}.
\end{enumerate}
While the use of sound activity mask has been previously noted in sequence modeling literature \cite{goodfellow2016deep}, it has not been explicitly addressed in the context of generative audio clustering.

\section{Experiments}
In this section we describe the experiment configuration for the different datasets, what metrics we use to evaluate clustering performance, the baselines, and results.

\subsection{Configuration} 

We set the latent space dimension with $d_\mathbf{z}=10$, which has shown to be sufficient in preliminary tests. Also, we opt for the lowest complexity operation with $M=1$ in the Monte Carlo sampling from \eqref{eq:carlo}. As previously described, the prior distribution over the latent space $p_\varphi(\mathbf{z})$ is modeled as a GMM. In our experiments, we use $K=10$ components in the mixture, matching the number of classes in the dataset. Furthermore, to adapt the model architecture to audio clustering, linear layers are used in the latent space to map the number of channels to the chosen latent dimension. This is a necessary modification compared to the fully connected VAE. The first linear layer transforms the encoder output channels to $2 \cdot d_\mathbf{z}$, where the first $d_\mathbf{z}$ elements represent the mean of the latent distribution, and the last $d_\mathbf{z}$ the variance. The second linear layer maps the latent representation to the decoder input dimension.

For the GMM prior $p_\phi(\mathbf{z})$, defined in \eqref{eq:gmmprior}, the initialization is as follows: i) the weighting vector $\pi_k$ is drawn from a uniform distribution with lower and upper bounds of 0.0 and 1.0, respectively; ii) the mean vectors $\boldsymbol{\mu}_k$ are initialized using Xavier uniform initialization; and iii) the covariance matrices $\boldsymbol{\Sigma}_k$ are initialized to zero. Note that the zero initialization of the covariance matrices implies initially deterministic components, which are updated during training. Neural network layers are initialized with the Kaiming uniform initialization for the weights and zero for the biases.

Moreover, we consider two settings. First, we apply the clustering methods to a dataset where the labels are not a complete abstraction, but an important feature of the data. For this case, we look into accuracy metrics as well as clustering metrics, as they should be synergetic and a good clustering should also mean a good classification. Still, time-frequency overlap is present to some extent, which should early-saturate the accuracy metrics. Second, we consider datasets where the labels are merely a human abstraction with low correlation with the actual time-frequency data features. This has been previously observed in literature \cite{shah2025bias,smart2024discipline}, and is tied to the societal, cultural, and institutional contexts or biases in which labels are constructed. This is common in background noise-like datasets, where the labels stand for where the recordings were made (airport, metro station, etc), and the time-frequency content often strongly overlaps between classes which share common acoustic properties \cite{heo2019acoustic}. Therefore, knowing that clustering quality is untied from the classification task, we focus on clustering metrics, leaving accuracy-relate metrics aside of the main analysis. Instead, we present the clustering metrics obtained with the actual labels from the dataset. The data considered for both tasks are described next.

\subsubsection{Spoken digits}
\label{sssec:spokendigits}
Spoken digit recognition consists of identifying which digit was spoken in an audio utterance. The main acoustic feature is the digit itself, while other sound characteristics are minor features. For this task, we consider the AudioMNIST dataset \cite{becker2023audiomnist}, with 30000 audio samples -- of which 24000 are randomly selected for training, 3000 for validation, and 3000 for testing -- where each file contains the audio recording, resampled at 16 kHz, of a spoken digit. The speakers are of different gender and age. The raw audio data are preprocessed as follows. First, we pad zeros to each audio sample until the desired duration of 1 second is achieved. The padded audio is then applied to a short-term Fourier transform (STFT), with length of 960 samples, Hann window of the same size, and a hop of 480 samples. Moreover, we take the module of the output of the STFT and limit the frequency range to 128 frequency bins -- approximately 6 kHz. Such a frequency range has showed to be sufficient for classification tasks on AudioMNIST in previous tests \cite{young2025hybridrealcomplexvaluedneural}. Finally, the spectrograms are normalized by their mean and variance, with ranges limited from 0 to 1 by a min-max adjustment.

The input to the neural network is the padded and preprocessed full second of audio, with 128 frequency bins and 99 time bins, which zero-frequency bin is removed. Additionally, we leverage time dependencies by feeding the entire duration of a file to the NN \cite{fiorio2024spectral}. Importantly, we generate an activity detection vector $\mathbf{a}_\mathbf{x}$ containing zeros in the zero-padded time-indexes, and ones otherwise. This vector multiplies both target and prediction during the calculation of the reconstruction loss, as in \eqref{eq:variationallowerboundclustering} and \eqref{eq:elbo}. 

\subsubsection{Acoustic environments} We devise the classification of urban acoustic scenes using real-world datasets. Such a task is much more challenging as the (background) sound features resemble noise and greatly overlap in time and frequency. We consider two different datasets: TAU2019 \cite{mesaros2018tau}, with 1200 audio recording files of 10 seconds from different acoustic scenes; and the UrbanSound8K (US8K) \cite{salamon2014us8k} dataset, with 8732 sound excerpts of 10 different acoustic scenes, mostly 4-seconds long. We use a mel-frequency cepstrum to reduce the input dimension without limiting frequency range, as acoustic scene classification can benefit from the broader range.

We resample data to 16 kHz. The US8K audio files are zero-padded to four seconds, where we also employ a sound activity mask for the calculation of the reconstruction error. The mask $\mathbf{a}_\mathbf{x}$ is always one for the TAU2019 dataset since all files are 10-seconds long, not requiring zero-padding. We obtain an STFT of 960 samples with a Hann window the same size and 50\% overlap. Lastly, we apply a mel-frequency scaling with 128 bins. The cepstrum is normalized by its mean and variance, and a min-max normalization to limit values from 0 to 1. We feed the NNs with a time-context window \cite{fiorio2024spectral} of 4 seconds for the US8K dataset and 10 seconds for the TAU2019 dataset, which are the maximum duration of the dataset's files.

Differently from the AudioMNIST case, urban scene classification has direct application for hearing devices \cite{yook2015environmentadaptive, lamarche2010adaptive}, where different acoustic scenes result in different processing, which is proportional to the constraints of the device. We take two cases into account: the same number of clusters as labels in the dataset; and a reduced number of clusters. Specifically for the considered urban acoustic scene datasets, we consider 10 clusters for the higher end, as it matches the number of labels in the dataset. For the lower end, we take 5 clusters into account, as it is a significant reduction from 10, merging similar clusters, but still sufficient for effectively calculating clustering metrics. In practice, a higher number of clusters results in a more complex and ``specialized'' processing, representing a higher-end version of, e.g., a hearing aid device. On the other hand, the lower cluster number could represent a more affordable version of the same device.

\subsection{Metrics}
\label{ssec:metrics}

The considered accuracy metrics, used for analysis in part 1, and the clustering metrics, taken into account for both parts (1 and 2) of the results, are described in the following.

\subsubsection{Unsupervised accuracy}
In (unsupervised) clustering tasks, the numeric labels may not correspond directly to the ground truth labels. We then consider an unsupervised approach for calculating accuracy, which consists of finding the matching truth labels for the clusters via the Hungarian algorithm \cite{kuhn1955hungarian}. Unsupervised accuracy ranges from 0 to 100\%.

\subsubsection{Normalized mutual information}
The normalized mutual information (NMI) is an information theoretic approach that evaluates the clustering quality by measuring the amount of shared information between clustering assignments and truth labels \cite{vinh2010information}. Its range is from 0 to 1.

\subsubsection{Silhouette score}
The Silhouette score \cite{rousseeuw1987silhouettes} measures how similar a data point is to its own cluster in comparison to other clusters. It combines cohesion (how close data points within a cluster are) and separation (how distinct is a cluster from another). The range is from -1 to +1: -1 indicates misclassification; 0 tells us that clusters overlap; and +1 indicate optimal clustering.

\subsubsection{Davies-Bouldin index}
The Davies-Bouldin index (DBI) \cite{davies1979cluster} is defined as the average similarity ratio of each cluster with the cluster that is most similar to it. A lower DBI indicates better clustering -- in terms of compactness and separation. Its range can vary from 0 to infinity.

\subsubsection{Cali\'nski-Harabasz index}
The Cali\'nski-Harabasz index (CHI) \cite{calinski1974dendrite} measures the ratio of the sum of between-cluster to within-cluster dispersion -- distinctiveness. Better-defined clusters are indicated by a higher value, and the range is from 0 to infinity.

\subsection{Baselines}
\label{ssec:results}

As benchmarks, we use two traditional methods, named K-means \cite{hartigan1979kmeans} and the optimization of a Gaussian mixture model using the expectation-maximization (EM) algorithm (GMM-EM) \cite{dempster1977em}. Other classical approaches and derivations are assumed to achieve similar performance to K-means and GMM-EM. 

We also compare our proposed M2-AC model directly to the VIB-GMM \cite{ugur2020variational} with its lower-sized alternative VIB-GMM-AC. Additionally, as noticed in \cite{ugur2020variational}, the Variational Deep Embedding (VaDE) \cite{jiang2017variational} is a subset of the VIB-GMM model. Therefore, for comparison, we also consider a model trained with the VaDE ELBO \cite{jiang2017variational}, also including a sound activity mask for audio clustering:
\begin{equation}
\label{eq:vade_elbo_compact}
\mathcal{L}_{\mathrm{VaDE}}(\theta,\phi) = 
\mathbb{E}_{q_{\phi}(\mathbf{z},c|\mathbf{x})}
[\mathbf{a}_{\mathbf{x}} \log p_{\theta}(\mathbf{x}|\mathbf{z})] - D_{KL}(q_{\phi}(\mathbf{z},c|\mathbf{x}) || p_{\theta}(\mathbf{z}| c).
\end{equation}
Equations \eqref{eq:xunxo_kl}, \eqref{eq:pc_gmm}, and \eqref{eq:k_argmax} can be adapted to the VaDE ELBO in (9) by replacing $p_\phi(\mathbf{z})$ with $p_\theta(\mathbf{z}|c)$ and $q_\phi(\mathbf{z}|\mathbf{x})$ with $q_\phi(\mathbf{z},c|\mathbf{x})$. The VaDE model uses the same architecture of the VIB-GMM model. Thus, we also consider its audio clustering version (VaDE-AC) as the one from Fig.~\ref{fig:VAE}.

Other variations in literature are often a derivation of the VaDE and the VIB-GMM, and they often are unable to outperform the considered methods consistently \cite{guo2025gammaclustering}. For all considered models, we use the same encoder and decoder architectures, detailed in Section~\ref{sec:architecture}, which include the modifications proposed in Section~\ref{ssec:practical}. All models are trained 10 times, independently, for 500 epochs, with a learning rate decreasing exponentially from 5e-4 to 5e-5, using Adam optimizer. In the following, we present the results for the spoken digits clustering task.

\subsection{Performance}

\begin{table}[!t]
    \centering
    \resizebox{\columnwidth}{!}{%
    \begin{tabular}{c c c c c c c c c}
    \hline
         & & & & & & \multicolumn{2}{c}{\textbf{Parameters}} \\
        \textbf{Method} & \textbf{Accuracy} (\%) \textcolor{darkgray}{$\uparrow$} & \textbf{NMI} \textcolor{darkgray}{$\uparrow$} & \textbf{Silhouette} \textcolor{darkgray}{$\uparrow$} & \textbf{DBI} \textcolor{darkgray}{$\downarrow$} & \textbf{CHI} $\times10^{3}$ \textcolor{darkgray}{$\uparrow$} & \textbf{Total} (M) & \textbf{Enc.} (M) \\
    \hline        
        \textcolor{gray}{None (labels)} & \textcolor{gray}{100.00} & \textcolor{gray}{1.00} & \textcolor{gray}{-0.04} & \textcolor{gray}{5.56} & \textcolor{gray}{0.10} & NA & NA \\
        K-means & 18.40 ± 1.22 & 0.10 ± 0.02 & 0.13 ± 0.01 & 2.04 ± 0.02 & 0.69 ± 0.07 & NA & NA \\
        GMM-EM & 17.62 ± 0.33 & 0.09 ± 0.07 & 0.13 ± 0.01 & 1.95 ± 0.04 & 0.69 ± 0.03 & NA & NA \\
        VaDE & 69.07 ± 7.90 & 0.76 ± 4.41 & 0.23 ± 0.16 & 1.65 ± 0.16 & 6.37 ± 41.78 & 15.26 & 7.63 \\
        VaDE-AC & 77.98 ± 9.11 & \textbf{0.81} ± 0.05 & 0.25 ± 0.02 & 1.51 ± 0.10 & 0.61 ± 0.05 & 2.07 & 1.31 \\
        VIB-GMM & \textbf{78.26} ± 5.08 & 0.78 ± 4.37 & 0.23 ± 0.02 & 1.56 ± 0.10 & 0.64 ± 0.03 & 15.26 & 7.63 \\
        VIB-GMM-AC & 70.78 ± 2.80 & 0.71 ± 0.02 & 0.21 ± 0.01 & 1.61 ± 0.26 & 0.54 ± 0.03 & 2.00 & 1.25 \\
        M2-AC ($\beta=0.5$) & 76.30 ± 7.58 & 0.78 ± 0.05 & \textbf{\textcolor{black}{0.97}} ± 0.01 & \textbf{\textcolor{black}{0.07}} ± 0.01 & \textbf{\textcolor{black}{40.14}} ± 10.96 & 3.32 & 1.24 \\
    \hline
    \end{tabular}
    }
    \caption{AudioMNIST dataset results, either by considering labels as clusters or by applying K-means, GMM-EM, and the aforementioned variational autoencoder models. The best results are highlighted in bold.}
    \label{tab:results_partA}
    \vspace{-4mm}
\end{table}

\subsubsection{Spoken digits} For spoken digit clustering, which results are shown in Table~\ref{tab:results_partA}, K-means and GMM-EM achieve insufficient accuracy and NMI metrics, only enhancing clustering scores when compared to the truth labels.

The variational autoencoder approaches are successful in achieving a sufficiently high accuracy (average of 74.5\%) and NMI (average of 0.77). The method achieving the highest accuracy is the VIB-GMM with 78.26\%, with its low-complexity counterpart, the VIB-GMM-AC, arriving at 70.78\%. The difference in parameters is, however, significant, with a reduction of 13.26 M parameters from the original VIB-GMM model. On another hand, the VaDE of higher complexity (15.26 M parameters) achieves lower accuracy, of 69.07\%, when compared to the hardware-friendly version (2.07 M parameters), with 77.98\% accuracy. It appears that the VaDE's less general loss function was more efficiently optimized by the audio clustering variant, with convolutions and recurrent layers instead of fully connected layers. The different behavior from VIB-GMM and VaDE versions seems counter intuitive, showing a strong dependence of loss function and neural network architecture. The silhouette, DBI, and CHI scores, nonetheless, are not very satisfactory for the VaDE and VIB-GMM variations. The consistent failure of the Gaussian mixture-like latent space in achieving high clustering metrics gives us a hint that the choice might not be appropriate for the task when more challenging scenarios are considered.

The M2-AC, on the other hand, achieves high accuracy and NMI (76.30\% and 0.78, respectively), and is capable of maintaining very high clustering metrics, with the silhouette metric almost at its maximum and lowering the DBI to almost zero, as well as achieving a very high CHI score. This indicates that the choice of a categorical latent space for clustering better aligns to the nature of the task, which is inherent categorical. Interestingly, no method was able to achieve an accuracy higher than 80.00\%. This result reflects on the time and frequency overlap of the spoken digits: even though the label of the digit is the principal feature, their physical overlap (of sound) provides a natural limit to perfect unsupervised classification. With those insights in mind, we analyze next the more challenging task of environment sound clustering.

\subsubsection{Acoustic environments} For this challenging task, GMM-EM did not converge when the TAU2019 or UrbanSound8K data were considered. The VaDE and its lower-complexity variation, the VaDE-AC, consistently presented divergence or latent space collapse\footnote{We refer to latent space collapse when all data is classified to a single cluster.}, even for different random seeds. 

First and foremost, as we can see in Table~\ref{tab:results_partB} how the labels from the datasets represent very low clustering scores, stemming from the disconnection of the labels themselves (location where the scenes were recorded) to their actual audio content. In terms of clustering, K-means was able to converge but achieved very poor clustering metrics. Moreover, the VIB-GMM-AC could process the data without any divergence or latent collapse, however, its higher-complexity format, the VIB-GMM, often would result in a collapse of the clusters. Indifferently, we can suppose that the problem for the VaDE probably stems from the loss function, given that VIB-GMM-AC and VaDE-AC have a very similar neural network architecture, but different losses, with the additional considerations in the VaDE ELBO \eqref{eq:vade_elbo_compact} becoming too strict of a constraint when applied to high-dimensional highly-overlapping data. Furthermore, by removing the VaDE loss function constraint, the general form considered for the VIB-GMM-AC was able to hold a stable training. Nevertheless, the metrics show unsuccessful clustering. Such scores are in line with expectations from the previous experiment, where we pointed out that the Gaussian mixture latent space is not the most interesting choice for clustering given its continuous nature, while clustering is, in essence, a categorical problem.

\begin{table}[!t]
    \centering
    \resizebox{\columnwidth}{!}{%
    \begin{tabular}{c c c c c c c c}
    \hline
         & & & & & & \multicolumn{2}{c}{\textbf{Parameters}} \\
        \textbf{Dataset} & \textbf{Clusters} & \textbf{Method} & \textbf{Silhouette} \textcolor{darkgray}{$\uparrow$} & \textbf{DBI} \textcolor{darkgray}{$\downarrow$} & \textbf{CHI} $\times10^{3}$ \textcolor{darkgray}{$\uparrow$} & \textbf{Total} (M) & \textbf{Enc.} (M) \\
    \hline
        \multirow{9}{*}{TAU2019} & \multirow{5}{*}{10} & \textcolor{gray}{None (labels)} & \textcolor{gray}{-0.05} & \textcolor{gray}{9.17} & \textcolor{gray}{0.11} & \textcolor{gray}{NA} & \textcolor{gray}{NA} \\
         & & K-means & 0.03 ± 0.01 & 3.48 ± 0.03 & 0.00 ± 0.00 & NA & NA \\
        &  & VIB-GMM-AC & -0.02 ± 0.02 & 5.07 ± 0.27 & 0.25 ± 0.01 & 2.20 & 1.28 \\
        &  & M2-AC ($\beta=2.0$) & \textbf{\textcolor{black}{0.77}} ± 0.01 & \textbf{\textcolor{black}{0.33}} ± 0.02 & \textbf{\textcolor{black}{6.97}} ± 0.38 & 3.67 & 1.26 \\
        &  & M2-AC-w ($\beta=2.0$) & \textbf{0.53 }± 0.01 & \textbf{0.61} ± 0.01 & \textbf{2.03} ± 0.05 & 3.27 & 1.24 \\
    \cline{2-8}
        & \multirow{4}{*}{5} & K-means & 0.06 ± 0.01 & 3.12 ± 0.06 & 0.53 ± 0.00 & NA & NA \\
        &  & VIB-GMM-AC & 0.02 ± 0.00 & 3.41 ± 0.14 & 0.34 ± 0.03 & 2.20 & 1.28 \\
        &  & M2-AC ($\beta=2.0$) & \textbf{\textcolor{black}{0.79}} ± 0.06 & \textbf{\textcolor{black}{0.30}} ± 0.07 & \textbf{\textcolor{black}{14.76}} ± 4.01 & 3.55 & 1.24 \\  
        &  & M2-AC-w ($\beta=2.0$) & \textbf{0.59} ± 0.01 & \textbf{0.54} ± 0.01 & \textbf{4.98} ± 0.10 & 3.24 & 1.23 \\
    \hline
        \multirow{9}{*}{UrbanSound8K} & \multirow{5}{*}{10} & \textcolor{gray}{None (labels)} & \textcolor{gray}{-0.06} & \textcolor{gray}{4.96} & \textcolor{gray}{0.04} & \textcolor{gray}{NA} & \textcolor{gray}{NA} \\    
         & & K-means & 0.14 ± 0.02 & 1.90 ± 0.14 & 0.20 ± 0.02 & NA & NA \\
        &  & VIB-GMM-AC & 0.09 ± 0.01 & 2.35 ± 0.21 & 0.11 ± 0.02 & 2.03 & 1.25 \\
        &  & M2-AC ($\beta=2.0$) & \textbf{\textcolor{black}{0.73}} ± 0.11 & \textbf{\textcolor{black}{0.39}} ± 0.14 & \textbf{\textcolor{black}{1.17}} ± 0.48 & 3.37 & 1.24 \\  
        &  & M2-AC-w ($\beta=2.0$) & \textbf{0.55} ± 0.01 & \textbf{0.61} ± 0.02 & \textbf{0.44} ± 0.03 & 3.27 & 1.24 \\
    \cline{2-8}
        & \multirow{4}{*}{5} & K-means & 0.18 ± 0.01 & 1.63 ± 0.10 & 0.32 ± 0.03 & NA & NA \\
        &  & VIB-GMM-AC & 0.10 ± 0.02 & 2.33 ± 0.34 & 0.14 ± 0.03 & 2.03 & 1.25 \\
        &  & M2-AC ($\beta=2.0$) & \textbf{\textcolor{black}{0.78}} ± 0.09 & \textbf{\textcolor{black}{0.32}} ± 0.10 & \textbf{\textcolor{black}{3.19}} ± 1.47 & 3.32 & 1.24 \\
        &  & M2-AC-w ($\beta=2.0$) & \textbf{0.61} ± 0.01 & \textbf{0.52} ± 0.01 & \textbf{1.10} ± 0.07 & 3.24 & 1.23 \\
    \hline
    \end{tabular}
    }
    \caption{Clustering metrics on the test set of the mentioned datasets, by applying K-means and the aforementioned variational autoencoder models, averaged over 10 independent runs. The best results are highlighted in bold.}    
    \label{tab:results_partB}
    \vspace{-4mm}
\end{table}

By considering a categorical latent space for clustering, the M2-AC model was able to achieve satisfactory silhouette, DBI, and CHI scores for both TAU2019 and UrbanSound8K datasets. Considering windows of 1 second of audio, we also applied the time-context windowing approach proposed in Section~\ref{sec:windowing} resulting in the M2-AC-w variation. The windowing approach reduced even further the complexity of the model with an obvious penalization in all clustering metrics, which is expected given that the new estimation approximates the entire signal by separate windows, averaging each output. The windowing approach is still, however, able to perform successful clustering for both considered datasets.

For a closer connection to practical applications, we freely modify the number of clusters to 5 and replicate the results for both datasets. This could represent a more affordable system, for example, where similar environments are grouped together and processed by a similar algorithm. From the results, we can notice that silhouette and DBI metrics are increased by small amounts, indicating that clusters are almost as compact and separate as before. A greater increase is perceived for the CHI score, which tells us that the lower number of clusters results in less disperse clusters, probably due to the reduction of outliers.

\begin{figure}[!t]
\centering
    \begin{subfigure}{.3\textwidth}
        \centering
        \includegraphics[width=0.9\textwidth]{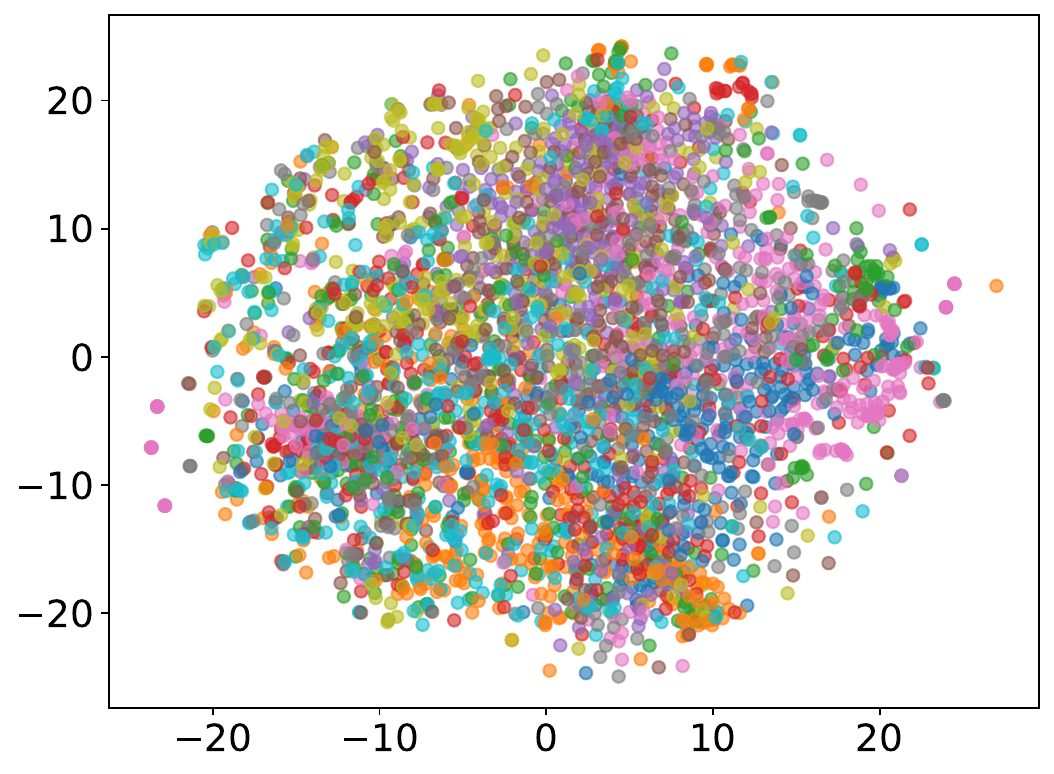}  
        \caption{TAU2019 test data}
        \label{fig:tau_data}
    \end{subfigure}
    \begin{subfigure}{.3\textwidth}
        \centering
        \includegraphics[width=0.9\textwidth]{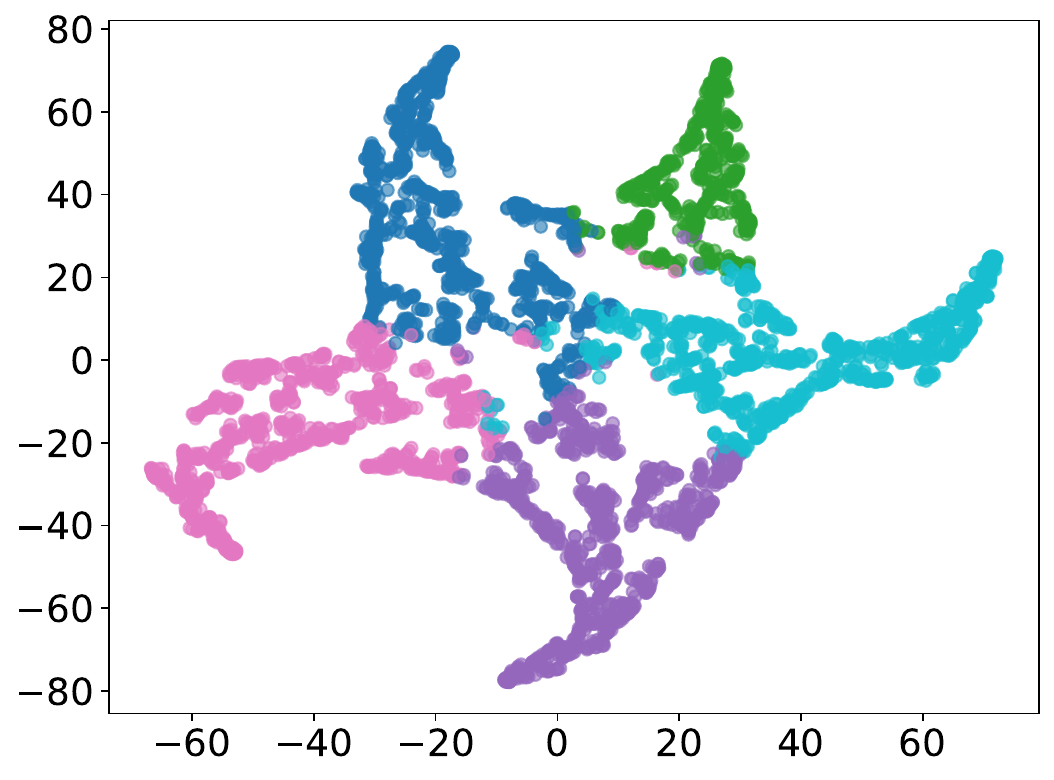}
        \caption{5 clusters}
        \label{fig:tau5_clusters}
    \end{subfigure} 
    \begin{subfigure}{.35\textwidth}
        \centering
        \includegraphics[width=0.9\textwidth]{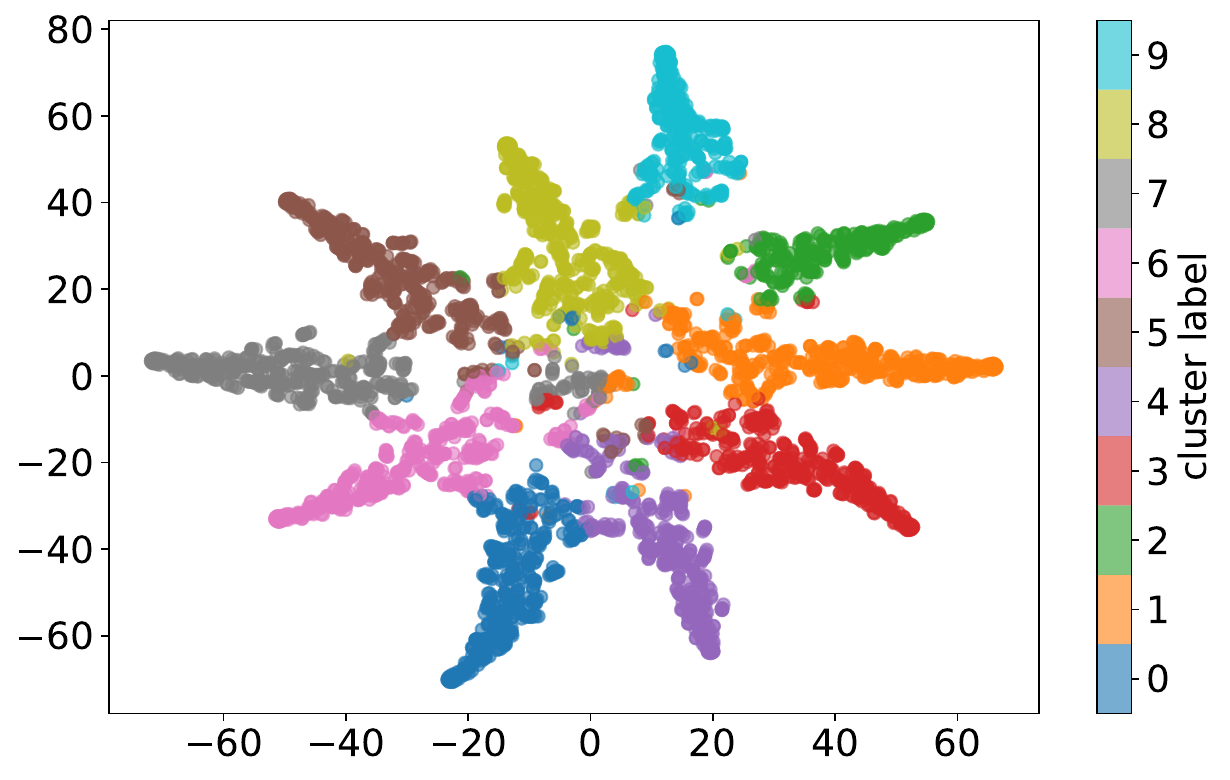}
        \caption{10 clusters}
        \label{fig:tau10_clusters}
    \end{subfigure} \\ \vspace{2mm}
    \begin{subfigure}{.3\textwidth}
        \centering
        \includegraphics[width=0.9\textwidth]{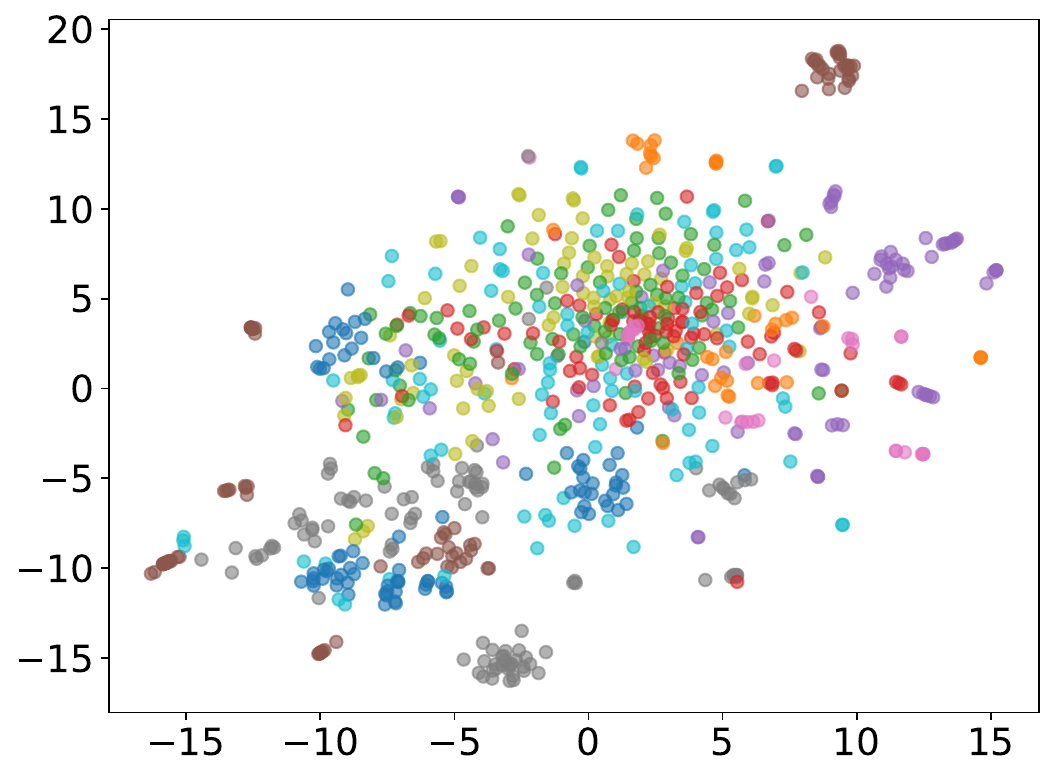}  
        \caption{US8K fold 1 data}
        \label{fig:us8k_data}
    \end{subfigure}
    \begin{subfigure}{.3\textwidth}
        \centering
        \includegraphics[width=0.9\textwidth]{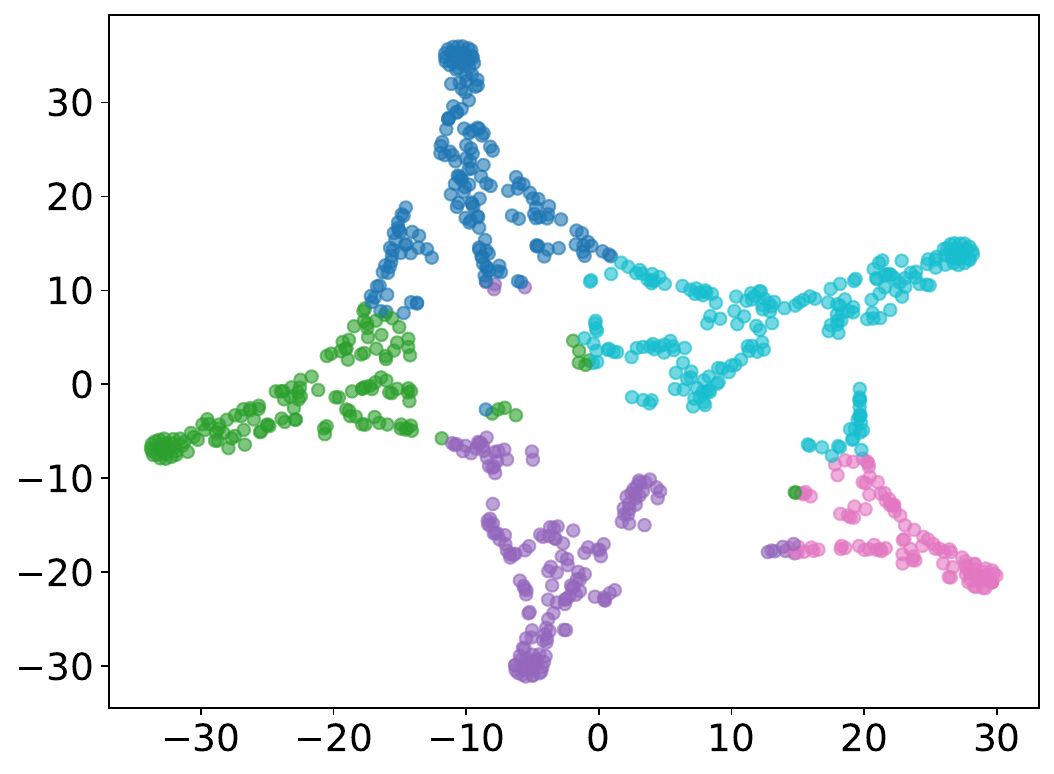}
        \caption{5 clusters}
        \label{fig:us8k5_clusters}
    \end{subfigure} 
    \begin{subfigure}{.35\textwidth}
        \centering
        \includegraphics[width=0.9\textwidth]{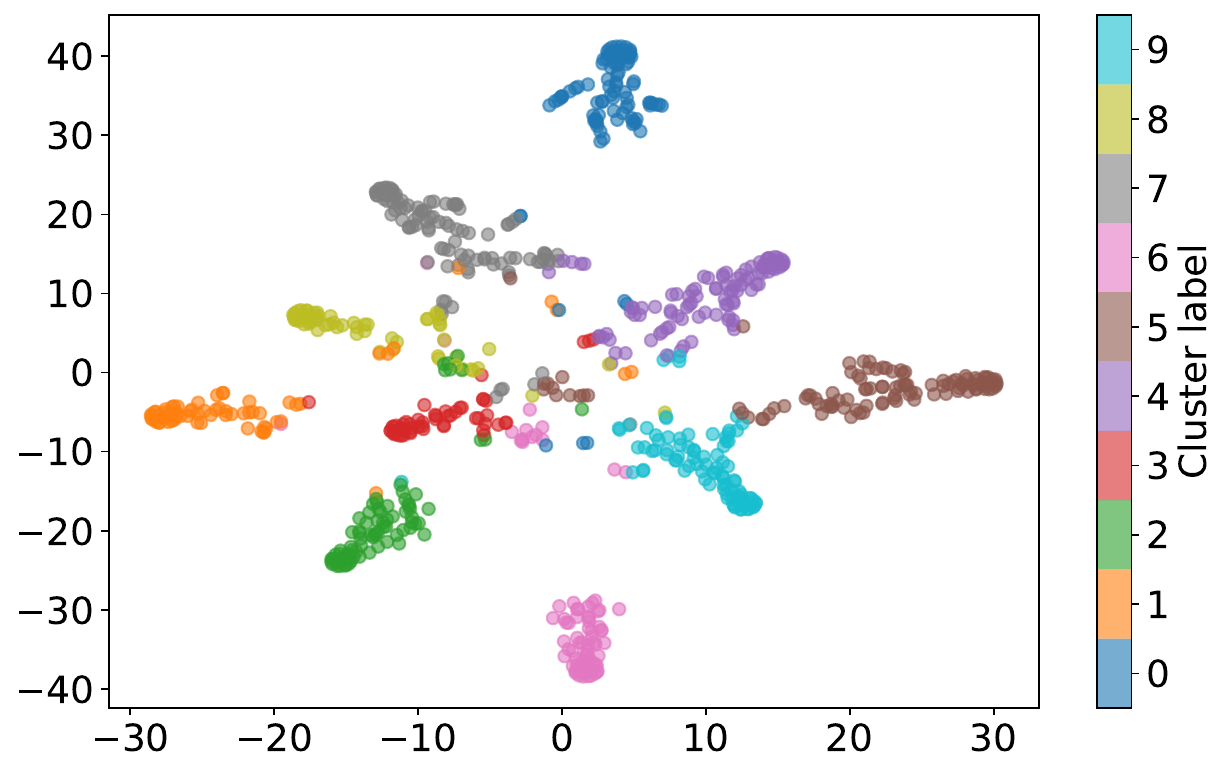}
        \caption{10 clusters}
        \label{fig:us8k10_clusters}
    \end{subfigure}
\caption{TAU2019 test set and US8K fold 1 raw and clustered data points by the M2-AC-w approach. Data size reduced for plotting with t-SNE. Each color represents a cluster (labels shown in bar plot), and each circle in the plot is a data point.}
\label{fig:clusters}
\vspace{-5mm}
\end{figure}

For a comprehensive visualization of the clustering process, we present an example of the raw and clustered data points of both datasets through the M2-AC-w approach in Fig.~\ref{fig:clusters}, obtained through t-distributed stochastic neighbor embedding (t-SNE). Note how the raw data have strongly-overlapped data points, which are nicely split apart in different clusters after the method is applied. Notice, also, how outliers are more present for 10 clusters than for 5 clusters, which explains the bigger leap in the CHI metric on the quantitative results. Importantly, there is always overlap, which happens given the time and frequency overlap of sound data. To achieve close-to-perfect classification given its abstract labels, one would have to rely on supervised learning, which would focus solely on that task and would less reflect the physical behavior of audio signals.

\section{Conclusion}

We proposed a clustering model tailored for audio applications based on an existing continuous-categorical variational inference process. The proposed method uses a convolutional-recurrent architecture that allows for substantial reduction in parameters. For a spoken digits clustering task, all considered variational approaches achieved high accuracy metrics, however, the proposed model achieve much higher clustering metrics, given its categorical nature, aligned with the clustering problem. Differently, the only method to succeed for clustering acoustic environment recordings was the proposed M2-AC and its windowed variation, the M2-AC-w. This indicates that the Gaussian mixture latent representation, often used when clustering images and text with a VAE architecture, is not enough for clustering strongly overlapped audio data. The latter requires the presence of a categorical latent space, which can achieve satisfactory clustering performance.

An unsupervised clustering system for acoustic environments can be used in hearing devices to cluster different soundscapes, based on the audio information. In practice, a sound input that is assigned to a cluster can be processed differently, with algorithms that are tailored to that type of data. This can be further explored as future work. We also suggest the modification of the current system to sound source localization/tracking, such that the latent space represents sampled positions of one or more sources.

\section{Acknowledgments}
This work was supported by the Robust AI for SafE (radar) signal processing (RAISE) collaboration framework between Eindhoven University of Technology and NXP Semiconductors, including a Privaat-Publieke Samenwerkingen toeslag (PPS) supplement from the Dutch Ministry of Economic Affairs and Climate Policy.

Moreover, the authors would like to thank Alex Young, Bruno Defraene, Frans Widdershoven, Johan David, and Yan Wu for the insightful comments during the development of this work. Additional thanks go to José Núñez Kasaneva for the discussion that planted the initial seed for the use of variational autoencoders.


\appendix

\section{Ablation studies}
\label{appendix:ablation_studies}

For the ablation study on the practical implementation to audio clustering (Section~\ref{ssec:practical}), we consider the problem of clustering of spoken digits with the AudioMNIST dataset, as described in detail in Section~\ref{sssec:spokendigits}. We utilize the M2-AC model architecture mentioned in Section~\ref{sec:architecture}. The models are trained for 500 epochs each, 10 times (otherwise specified), independently. The Adam optimizer is considered, with a decreasing learning rate from 5e-4 to 5e-5. The model without recurrent layers is defined by removing all GRU layers from the architecure shown in Fig.~\ref{fig:VAE-M2}. In the case of 1D convolutions, all Conv2D and BatchNorm2D are modified to a single dimension operating only in the frequency axis, with kernel 8, stride 2, and padding 3, maintaining the same output channel values as indicated in the schematic.

The ablation study results in Table~\ref{tab:ablation_cuvac}, with metrics described in Section~\ref{ssec:metrics}, show that the utilization of 2D convolutions and recurrent layers in synergy achieves the best clustering results, as well as high unsupervised accuracy for the spoken digits when compared to the dataset labels. Notice that removing recurrent layers results in a much higher standard deviation in terms of accuracy and NMI. That reflects how a different initialization can strongly affect the results of the model, showing that recurrent layers add substantial robustness to the process. Additionally, the inclusion of a sound activity mask in the ELBO \eqref{eq:elbo} is crucial for the correct clustering, otherwise, the model is unable to focus solely on the sound-rich part of the audio file, sometimes even clustering zero-padding patterns as we observed by looking at individual results, i.e., the clustering metrics without activity mask cannot be trusted for zero-padded data. Importantly, three independent runs had their latent space collapsed when the mask was removed, and all data was clustered into a single cluster. 

\begin{table}[!t]
    \centering
    \resizebox{\columnwidth}{!}{%
    \begin{tabular}{c c c c c c c c}
    \hline
        \textbf{Conv.} & \textbf{Rec} & \textbf{Act.} & & & & & \\
        \textbf{type} & \textbf{layers} & \textbf{mask} & \textbf{Acc.} (\%) \textcolor{darkgray}{$\uparrow$} & \textbf{NMI} \textcolor{darkgray}{$\uparrow$} & \textbf{Sil.} \textcolor{darkgray}{$\uparrow$} & \textbf{DBI} \textcolor{darkgray}{$\downarrow$} & \textbf{CHI} $\times10^{3}$ \textcolor{darkgray}{$\uparrow$} \\
    \hline        
        2D & Yes & Yes & \textbf{76.30} ± 7.58 & \textbf{0.78} ± 0.05 & \textbf{\textcolor{black}{0.97}} ± 0.01 & \textbf{\textcolor{black}{0.07}} ± 0.01 & \textbf{\textcolor{black}{40.14}} ± 10.96 \\
        1D & Yes & Yes & 22.32 ± 11.49 & 0.13 ± 0.18 & 0.90 ± 0.02 & 0.15 ± 0.03 & 12.42 ± 6.10 \\
        2D & No & Yes & 39.20 ± 21.46 & 0.38 ± 0.28 & 0.85 ± 0.12 & 0.25 ± 0.15 & 13.74 ± 11.27 \\
        2D & Yes & No & 52.69 ± 5.36 & 0.58 ± 0.64 & 0.94 ± 0.01 & 0.13 ± 0.06 & 21.13 ± 4.27 \\
    \hline
    \end{tabular} 
    }
    \caption{Clustering and accuracy metrics on the test set of AudioMNIST by applying the M2-AC model with different configurations related to the practical implementation to audio. Average over 10 independent runs with standard deviation. For the case without activity mask, only 7 runs were considered as the remaining 3 had their latent space collapsed.}
    \label{tab:ablation_cuvac}
\end{table}


 \bibliographystyle{elsarticle-num} 
 \bibliography{refs}






\end{document}